\documentclass[a4paper,11pt, draftcls, onecolumn]{IEEEtran}

\usepackage{graphics, graphicx}
\usepackage{amssymb}
\usepackage{amsmath}
\usepackage[rm]{subfigure}    
\usepackage{threeparttable}   
\usepackage[dvips]{color}
\usepackage{url}   
\usepackage[latin1]{inputenc} 
\usepackage[T1]{fontenc}
\usepackage{algorithm}              
\usepackage{algorithmic}
\usepackage{cite}
\usepackage{multirow}
\usepackage[dvips]{color}
\usepackage{lineno}
\usepackage[normalem]{ulem}
\usepackage{soul}  % use \st{works}, where works will be marked with a medium line across the works
\usepackage{algorithm}
\usepackage{indentfirst}
\usepackage{setspace}
\usepackage{cases}

\allowdisplaybreaks

\newcommand{\bm}[1]{\mbox{\boldmath{$#1$}}}

\newtheorem{theorem}{Theorem}
\newtheorem{lemma}[theorem]{Lemma}

\usepackage{tikz}
\definecolor{gold}{rgb}{0.85,.66,0}
\definecolor{cian}{rgb}{.02,.7,.95}
\definecolor{ppp}{rgb}{.7,.3,.82}
\definecolor{red}{rgb}{1,0,0}
\newcommand{\colk}{\textcolor{black}}

\newcommand{\cas}{\overset{a.s.}{=}}

\begin{document}
%--------------------------------------------------
\title{Pilot Distribution Optimization {and Power Control} in Multi-Cellular Large Scale MIMO Systems}
\author{José Carlos Marinello \and Taufik Abrão 
\thanks{The authors are with  the Communications Group, Department of Electrical Engineering, Londrina State University, Paraná, Brazil. Emails:  \text{zecarlos.ee@gmail.com}; \text{taufik@uel.br}; \quad \text{http://www.uel.br/pessoal/taufik}.}}
\maketitle

\begin{abstract}
Massive MIMO communication systems have been identified as one of the most prominent technologies of the next generation wireless standards, such as 5G, due to the large gains in energy and spectral efficiency that can be achieved. In the asymptotic condition of infinite number of antennas at the base station (BS), the performance bottleneck of these systems is due to the pilot contamination effect, \emph{i.e.}, the directional interference arising from users in adjacent cells that reuse the same set of orthogonal training sequences, and thus the interference seen by each user is determined by the pilot sequence assigned to him. We show in this paper that the system performance can be improved by appropriately assigning the pilot sequences to the users, in the so-called pilot allocation scheme. Depending on the optimization metric adopted, it is more advantageous to a user with certain long-term fading coefficient be assigned to a specific pilot sequence, whose interference can be completely estimated in advance by the BS by only knowing the long term fading coefficients of users in adjacent cells. Besides, if the objective is to maximize the number of users with a target quality of service, we have shown that the pilot allocation schemes can be combined with power control algorithms, resulting in much more improvements for the system. For unitary frequency reuse factor, we have found that the data throughput provided for 95\% of the users increases when applying power control algorithm from  134kbps to 1.461Mbps with no pilot allocation, while this performance gain provided by power control changes from 793kbps to 6.743Mbps when pilot allocation is employed. If the reuse factor increases to 3, a 95\%-likely data throughput of 17.310Mbps can be assured when pilot allocation and power control are suitably combined.\\ 
\textbf{Keywords}: Massive MIMO systems; BER, Multiuser MIMO; Precoding; Pilot sequences
\end{abstract}

\newpage
%\linenumbers
%========================
%%%%%%%%%%%%%%%%%%%%%%%%%
\section{Introduction}
%%%%%%%%%%%%%%%%%%%%%%%%%
%========================
{Recently, massive multiple-input-multiple-output (MIMO) has attracted increased research attention in wireless communication fields \cite{Hanzo15}. 
In order to fully exploit the benefits of conventional MIMO systems, this concept has been proposed by increasing the number of base station (BS) antennas $N$ to infinity \cite{Marzetta10}.  Due to its advantages of very high spectral/energy efficiencies and increased reliability \cite{Ngo13_EE_SE}, massive MIMO systems are viewed as a potential technology for physical layer in next telecommunications standards, such as 5G \cite{Heath14}.} 

It was shown in \cite{Marzetta10} that in a time division duplex (TDD) noncooperative multi-cell MIMO system, employing training pilots for channel state information (CSI) acquisition in the uplink and an infinite number of BS antennas, the effects of uncorrelated thermal noise and fast fading are averaged out. Hence, the only factor that remains limiting performance in the large MIMO scenario is inter-cell interference, that when associated with the finite time available to send pilot sequences makes the estimated CSI at one BS ``contaminated'' by the CSI of users in adjacent cells, in the so-called pilot contamination effect. This phenomenon  results from unavoidable reuse of reverse-link pilot sequences by terminals in different cells. As a consequence of increasing the number of BS antennas to infinity, the transmit power can be designed arbitrarily small, since interference decreases in the same rate of the desired signal power, \textit{i.e.}, signal-to-interference-plus-noise ratio (SINR) is independent of transmit power \cite{Marzetta10}.

Alternative strategies to achieve better CSI estimates exist, such as \textbf{a)} frequency division duplex (FDD) \cite{Choi13}, in which pilots for CSI acquisition are transmitted in downlink, and estimates are fed back to BS in a feedback channel; and  \textbf{b)} network MIMO \cite{Valenzuela06}, where CSI and information data of different coordinated cells are shared among them in a backhaul link, creating a distributed antenna array that serves the users altogether. However, both schemes become unfeasible when $N \rightarrow \infty$ \cite{Marzetta10}, since lengths of forward pilot sequences and capacity of backhaul links increase substantially with $N$. Therefore, TDD has been assumed in this work without CSI sharing among different cells.

Operating with a large excess of BS antennas compared with the number of terminals $K$ is a challenging but  desirable condition, since some results from random matrix theory become applicable \cite{Debbah13}. It is known, for instance, that very tall/wide matrices tend to be very well conditioned, since their singular values distribution appears to be deterministic, showing a stable behavior (low variances) and a relatively  narrow spread \cite{Rusek13}. Besides, in the large scale MIMO, the most simple reception/transmission techniques, \textit{i.e.}, maximum ratio combining (MRC) deployed in the uplink and the matched-filtering (MF) precoding used in the dowlink, become optimal \cite{Rusek13}. 

When considering the multi-cell environment, it is found in \cite{Rusek13} that the asymptotic SINR of MF outperforms that of zero-forcing (ZF), although MF requires much more antennas to approach the asymptotic condition. On the other hand, a more rigorous expression for the achievable SINR of MF precoding in massive MIMO systems, in comparison with that derived in \cite{Marzetta10} and adopted in \cite{Rusek13}, has been obtained in \cite{Fernandes13}. Authors of latter showed that, for downlink, the effect of the transmit power constraint at BS still accounts in the massive MIMO regime, as opposed to what was assumed in \cite{Marzetta10}. Besides, authors of \cite{Fernandes13} discuss an efficient technique for temporally distribute the uplink transmissions of pilot sequences, avoiding simultaneous transmissions from adjacent cells and reducing interference as well, in conjunction with power allocation strategy. 

An analysis of {\it non-linear precoding} techniques applied to the downlink of a massive MIMO system is conducted in \cite{Melzer15}. Authors investigated the time domain vector perturbation (TDVP) scheme, which has been shown previously to almost achieve the downlink capacity of conventional MIMO channels. However, in the large-system analysis, it was shown that linear precoding schemes outperforms TDVP, in terms of increased sum rates, regardless of the user scheduling method adopted. Thus, linear precoding techniques has been investigated in our contribution. In \cite{Hanzo_SPR_2015}, the pilot contamination is tackled by dividing the users within each cell in two groups, which are: the center users, and the edge users. As the edge users would suffer from severe pilot contamination if the same set of pilot sequences were reused by every cell, it is assigned for each edge user an exclusive training sequence in a cluster of $L$ cells, while the center users reuse the same pilot's set. Although this so-called soft pilot reuse scheme effectively reduces the pilot contamination, the cost of devoting orthogonal pilots for every edge user may limit its practical appeal in TDD systems. Hence, we aim to reduce the pilot contamination in this paper adopting the challenging but realistic scenario of full pilot reuse among cells.

{Power allocation is an efficient form of improving the performance of wireless communication networks. For the multicell massive MIMO scenario, the problem of power allocation was investigated in \cite{Zhang15} under the optimization metric of maximizing the sum rate per cell. It was shown that the proposed method achieved substantial gains over the equal power allocation assignment. As opposed to a joint (across all cells) optimization, authors proposed a much simpler scheduling method to plan the power allocation arrangements for different cells, achieving almost the same performance as the joint optimization. However, for a practical system, maximizing the sum rate per cell is not the most suitable optimization metric, since the performance of some users (tipically those at the edge) may be severely penalized in order to provide very increased rates for another ones. In this paper we have adopted a fairer metric, which consists of providing a target performance for the majority of the users. Thus, the algorithm of \cite{Rasti11}, that was also extended to massive MIMO systems in \cite{Fernandes13}, is very suitable in this context.}

In this paper, we focus on the pilot distribution optimization combined with power control and its impact on the performance of multi-cellular massive MIMO systems. The novelty and contributions of this paper are as follows:
\begin{itemize}
\item[i.]  Different from previous works that have analysed the SINR and the capacity of the massive MIMO system \cite{Marzetta10, Fernandes13, Rusek13}, we investigate also its downlink uncoded bit-error rate (BER) performance, which is another important figure of merit in communication systems. An exact expression for the BER of each user is derived, depending on the transmit power of users and on the long-term fading coefficients.
\item[ii.]  We propose a novel and expedite method of optimizing the massive MIMO downlink transmission under different metrics, based on our derived BER expression, and on the asymptotic SINR expression of \cite{Fernandes13}. This method consists of simply assigning the available training sequences among the users within a cell in an efficient manner, by knowing only the power and the long-term fading coefficients of users in adjacent cells that reuse such pilot sequences. Different pilot allocation metrics enable us to minimize the \colk{average} BER, or maximize the average SINR, minimize the maximal BER or even maximize the minimum SINR. The proposed algorithms can lead to appreciable performance gains, both in terms of data rate, as well as in terms of BER of a massive MIMO system.
{\item[iii.] The proposed pilot allocation method that maximizes the minimum SINR is combined with power control, and it is shown that more significant gains can be achieved by the power control algorithm when pilot allocation is employed.}
\end{itemize}

The paper is organized as follows. Beyond this introductory section, the system model and asymptotic limits of the massive MIMO system are revisited and extended {are developed in Section \ref{sec:model}}. Our proposed methods of assigning the pilots among the users within the cell in an efficient manner, namely the pilot allocation (PA) schemes, are presented in Section \ref{sec:PA_Schemes}, {while the power {allocation strategy is investigated} in Section \ref{sec:PowC}}. Representative numerical results are discussed in Section \ref{sec:results}, while Section \ref{sec:concl} concludes the paper.

{\textit{Notations:} Boldface lower and upper case symbols represent vectors and matrices, respectively. ${\bf I}_N$ denotes the identity matrix of size $N$, while ${\bf 1}_K$ and ${\bf 0}_K$ are the unitary vector and null vector of length $K$, respectively. The transpose and the Hermitian transpose operator are denoted by $\{\cdot\}^T$ and $\{\cdot\}^H$, respectively; ${\rm diag}(\cdot)$ is the diagonal matrix operator; $||\cdot||$ {holds for} Euclidean norm of a vector, and $\min[\cdot]$ returns the minimum element of the input set. We use $\mathcal{CN}({\bf m}, {\bf V})$ to refer to a circular symmetric complex Gaussian distribution with mean vector ${\bf m}$ and covariance matrix ${\bf V}$. Also, $\mathbb{E}[\cdot]$ denotes the expectation operator, ${\rm u}[x]$ is the Heaviside step function (${\rm u}[x] = 1$ if $x\geq0$, ${\rm u}[x] = 0$ otherwise), while $\delta_{ij}$ is the Kronecker delta function ($\delta_{ij} = 1$ if $i=j$ and 0 otherwise).}

%============================================
\section{System Model}\label{sec:model}
%============================================
\colk{The adopted MIMO system is composed by $L$ BSs, each equipped with $N$ transmit antennas, reusing the same spectrum and the same set of $K$ pilot signals. Since TDD is assumed, reciprocity holds, and thus CSI is acquired by means of uplink training sequences. During a channel coherence time interval, the symbol periods are divided in uplink pilot transmissions, processing, downlink and uplink data transmissions \cite{Fernandes13}, \cite{Hanzo_SPR_2015}. Using orthogonal pilot sequences, the number of available sequences is equal to its length, $K$. Thus, $K$ is limited due to mobility of the users, which reduces the coherence time of the channel. Orthogonal frequency-division multiplexing (OFDM) is assumed in the same way as in \cite{Marzetta10}. The channel coherence band is divided into $N_{\textrm{smooth}}$ subcarriers, and each subcarrier is shared by $K$ users in the training stage. As discussed in \cite{Marzetta10}, dividing the channel coherence band by the subcarrier spacing,
\begin{equation}\label{eq:Nsmooth}
N_{\textrm{smooth}} = \frac{1 - \Delta t_{CP}}{\Delta t_{CP}},
\end{equation}
where $\Delta t_{CP}$ is the fraction of the OFDM symbol duration devoted to the cyclic prefix, tipically about $7\%$ in current standards. Note that only one out of $N_{\textrm{smooth}}$ subcarriers is assigned to a certain user for each coherence band in the training stage; therefore, a total number of $K \cdot N_{\rm smooth}$ users is allowed for each cell. 
We assume perfect orthogonality in the frequency domain, such that interference is only due to the $K$ users sharing the same subcarrier. Hence, we define our model for a generic subcarrier, assuming flat fading environment in which the BS communicates with $K$ users equipped with single-antenna mobile terminals (MTs)}. We denote the $1 \times N$ channel vector between the $\ell$th BS and the $k$th user of $j$th cell by ${\bf g}_{\ell k j} = \sqrt{\beta_{\ell k j}} {\bf h}_{\ell k j}$, in which $\beta_{\ell k j}$ is the long-term fading power coefficient, that comprises path loss and log-normal shadowing, and ${\bf h}_{\ell k j}$ is the short-term fading channel vector, that follows ${\bf h}_{\ell k j} \sim \mathcal{CN}({\bf 0}_N, {\bf I}_N)$. The channel matrix $\bf{H}$ is admitted constant over the entire frame and changes independently from frame to frame (block fading channel assumption). Note that $\beta_{\ell k j}$ is assumed constant for all $N$ BS antennas. For the $k$th user of each cell in a given subcarrier, it is assigned the sequence ${\bm \psi}_k = [\psi_{1 k} \,\, \psi_{2 k} \ldots \psi_{K k}]^T$, ${\bm \psi}_k \in \mathbb{C}^{K\times1}$. It holds that $|\psi_{i\,k}| = 1$ and $|{\bm \psi}_{k}^H \,{\bm \psi}_{k'}| = K \delta_{kk'}$ since the set of sequences is orthogonal.

In the training transmission phase, we have assumed synchronization in the uplink pilot transmissions, since this situation characterizes the worst case for inter-cellular interference \cite{Marzetta10}. Hence, the $N \times K$ received signal at the $\ell$th BS is:
\begin{equation}\label{eq:rx_pilots}
{\bf Y}_{\ell} = \sum_{j = 1}^{L} {\bf G}_{\ell j}^T \sqrt{{\bm \Gamma}_j} {\bm \Psi} + {\bf N},
\end{equation}
where ${\bm \Gamma}_j = {\rm diag}(\gamma_{1 j}  \,\, \gamma_{2 j} \, \ldots \, \gamma_{K j})$, being $\gamma_{k j}$ the uplink transmit power of the $k$th user of $j$th cell, ${\bf G}_{\ell j} = [{\bf g}_{\ell 1 j}^T  \,\, {\bf g}_{\ell 2 j}^T \, \ldots \, {\bf g}_{\ell K j}^T]^T$, such that the $K\times N$ matrix ${\bf G}_{\ell j} = \sqrt{\mathcal{B}_{\ell j}} {\bf H}_{\ell j}$, $ \mathcal{B}_{\ell j} = {\rm diag}(\beta_{\ell 1 j}  \,\, \beta_{\ell 2 j} \, \ldots \, \beta_{\ell K j})$, ${\bf H}_{\ell j} = [{\bf h}_{\ell 1 j}^T  \,\, {\bf h}_{\ell 2 j}^T \, \ldots \, {\bf h}_{\ell K j}^T]^T$ is of dimension $K\times N$, ${\bm \Psi} = [{\bm \psi}_1  \,\, {\bm \psi}_2 \, \ldots \, {\bm \psi}_K]$, and ${\bf N}$ is a $N \times K$ additive white Gaussian noise (AWGN) matrix \colk{whose entries have} zero mean and unitary variance.  

In order to generate the estimated CSI matrix $\widehat{\bf G}_{\ell}$ of their served users, the $\ell$th BS correlates its received signal matrix with the known pilot sequences:
\begin{equation}\label{eq:rx_pilots_correlator}
\widehat{\bf G}_{\ell}^T = \frac{1}{K} {\bf Y}_{\ell} {\bm \Psi}^H = \sum_{j = 1}^{L} {\bf G}_{\ell j}^T \sqrt{{\bm \Gamma}_j} + {\bf N}',
\end{equation}
where ${\bf N}' \in \mathbb{C}^{N\times K}$ is an equivalent AWGN matrix with zero mean and variance $\frac{1}{K}$. Hence, the channel estimated by the $\ell$th BS is contaminated by the channel of users that use the same pilot sequence in all other cells.

Information transmit symbols of the $\ell$th cell is denoted by the $K\times1$ vector ${\bf x}_{\ell} = [x_{1 \ell}  \,\, x_{2 \ell} \, \ldots \, x_{K \ell}]^T$, where $x_{k \ell}$ is the transmit symbol to the $k$th user of the $\ell$th cell, and takes a value from the squared quadrature amplitude modulation ($M$-QAM) alphabet, \colk{normalized in order to preserve unitary average power}. 
For analysis simplicity, using matrix notation, the $K\times 1$ complex-valued signal received by users of the $\ell$th cell is written as:
\begin{equation}\label{eq:rx_C}
{\bf r}_{\ell} = \sum_{j=1}^{L} {\bf G}_{j \ell} {\bf P}_j \sqrt{{\bm \Phi}_j} {\bf x}_j + {\bf n}_{\ell},
\end{equation}
where ${\bm \Phi}_j = {\rm diag}(\phi_{1 j}  \,\, \phi_{2 j} \, \ldots \, \phi_{K j})$, being $\phi_{k j}$ the downlink transmit power devoted by the $j$th BS to its $k$th user, ${\bf P}_j$ denotes the complex valued $N \times K$ precoding matrix of the $j$th BS, being each column \colk{${\bf p}_{k j}$} the $N \times 1$ precoding vector of the $k$th user. Finally, ${\bf n}_{\ell} \sim \mathcal{CN}({\bf 0}_K, {\bf I}_K)$ represents the AWGN vector observed at the $K$ MTs of the ${\ell}$th cell. 

Under the matched-filter beamforming technique, the vector \colk{${\bf p}_{k j}$} is computed as \cite{Fernandes13}:
\begin{equation}\label{eq:MF_prec}
{\bf p}^\textsc{mf}_{k j} = \frac{{\bf \widehat{g}}_{j k}^H}{||{\bf \widehat{g}}_{j k}^H||} = \frac{{\bf \widehat{g}}_{j k}^H}{\alpha_{k j} \sqrt{N}},
\end{equation}
in which $\alpha_{k j} = \frac{||{\bf \widehat{g}}_{j k}^H||}{\sqrt{N}}$, and ${\bf \widehat{g}}_{j k}$ is the $k$th row of the matrix $\widehat{\bf G}_{j}$. Note that the normalization in \eqref{eq:MF_prec} is necessary to satisfy the maximum transmit power available at the BS. 

In the same way, in the zero-forcing beamforming technique, the vector \colk{${\bf p}_{k j}$} is computed as:
\begin{equation}\label{eq:ZF_prec}
{\bf p}^\textsc{zf}_{k j} = \frac{{\bf w}_{j k}}{||{\bf w}_{j k}||},
\end{equation}
in which the vector ${\bf w}_{j k} = \widehat{\bf G}_{j}^H {\bf a}_{j k}$, and ${\bf a}_{j k}$ is the $k$th column of ${\bf A}_{j} = \left[ \widehat{\bf G}_j \widehat{\bf G}_{j}^H \right]^{-1}$. 

%============================================
\subsection{Asymptotic Limits of Massive MIMO}\label{sec:Massive_MIMO}
%============================================
Most of the asymptotic limits for massive MIMO systems can be build upon the following well known lemma: 
\begin{lemma}\label{wlln}\it
Let ${\bf s}_1$,${\bf s}_2 \in \mathbb{C}^{N\times1}$ be two independent complex-valued vectors following a normal distribution, with zero mean and variance $\sigma^2$. Then
\begin{equation}
\lim_{N \rightarrow \infty} \frac{{\bf s}_1^H {\bf s}_2}{N} \cas 0 \quad \text{and} \quad \lim_{N \rightarrow \infty} \frac{{\bf s}_1^H {\bf s}_1}{N} \cas \sigma^2.
\end{equation}
\end{lemma}
Since the channel vectors of different users can be seen as independent random vectors, the above lemma is widely used for deriving limits in the massive MIMO scenarios. It can be justified since as the vector's length grows, the inner products between independent vectors grow at lesser rates than the inner products of vectors with themselves. 

From \eqref{eq:rx_pilots_correlator}, it is proved in \cite{Fernandes13} that $\alpha_{k j}^2 \cas \sum_{l=1}^{L} \gamma_{k l} \beta_{j k l} + \frac{1}{K}$. Then authors show that $r_{k \ell}$, \textit{i.e.}, the received signal at the $k$th user of $\ell$th cell, can be written as \cite[Eq. (5)]{Fernandes13}:
\begin{equation}\label{eq:rx_kl_mf}
r_{k \ell} = \sum_{l=1}^{L} \sum_{j=1}^{K} \sqrt{\phi_{j l} \beta_{l k \ell}} {\bf h}_{l k \ell}^H {\bf p}^\textsc{mf}_{j l} x_{j l} + n_{k \ell}.
\end{equation}

Based on \eqref{eq:MF_prec} and Lemma \ref{wlln}, {equation} \eqref{eq:rx_kl_mf} can be simplified when $N \rightarrow \infty$ as:
\begin{eqnarray}\label{eq:rx_kl_mf_simp}
r_{k \ell} &=& \sum_{l=1}^{L} \sqrt{\phi_{k l} \beta_{l k \ell}} {\bf h}_{l k \ell}^H {\bf p}^\textsc{mf}_{k l} x_{k l} + n_{k \ell}, \nonumber\\
&=& \sum_{l=1}^{L} \frac{1}{\alpha_{k l}} \sqrt{N \phi_{k l} \gamma_{k \ell}} \beta_{l k \ell} x_{k l} + n_{k \ell}, \nonumber\\
&=& \sqrt{N \gamma_{k \ell}} \sum_{l=1}^{L} \frac{\sqrt{\phi_{k l}} \beta_{l k \ell} x_{k l}}{\alpha_{k l}} + n_{k \ell}.
\end{eqnarray}
Note that the AWGN of the estimated CSI in \eqref{eq:rx_pilots_correlator} vanishes in \eqref{eq:rx_kl_mf_simp}. This occurs since it is independent of ${\bf h}_{l k \ell}^H$, and thus its product as $N \rightarrow \infty$ is averaged out according to Lemma \ref{wlln}.

From \eqref{eq:rx_kl_mf_simp}, it is straightforward to see the asymptotic downlink SINR of the system as:
\begin{eqnarray}\label{eq:ultimate_SINR}
\text{SINR}^\textsc{dl}_{k \ell} \, &=& \lim_{N \to \infty} \frac{N \gamma_{k \ell} \, \phi_{k \ell} \beta^2_{\ell k \ell}/\alpha^2_{k \ell}}{N \gamma_{k \ell}\, \left(\sum_{\substack{j = 1 \\ j \neq \ell}}^L  \phi_{k j} \beta^2_{j k \ell}/\alpha^2_{k j}\right) + 1} = \frac{\phi_{k \ell} \beta^2_{\ell k \ell}/\alpha^2_{k \ell}}{\sum_{\substack{j = 1 \\ j \neq \ell}}^L  \phi_{k j} \beta^2_{j k \ell}/\alpha^2_{k j}}.
\end{eqnarray}
\normalsize

Note that this limit depends mainly on the long-term fading coefficients $\beta_{j k i}$, which are related to the spatial distribution of the users in the different cells.

{It can be shown that, when the constraint of maximum transmit power available at BS is considered as in \eqref{eq:MF_prec} and \eqref{eq:ZF_prec}, both MF and ZF precoding schemes converges to the same precoding vector when $N \rightarrow \infty$. Therefore, the assymptotic limits for the massive MIMO system are valid for both {precoding} techniques. However, this equality holds only for $N$ very large. For intermediate values, it is seen that the ZF {precoding} scheme approaches the asymptotic limit faster than the MF beamforming, as numerically demonstrated in Section \ref{sec:results_conv}. By the way, the MF technique is quite less complex, and can be implemented in a decentralized way since the precoding vector of each user is not dependent on the estimated channels of other users, as opposed to ZF.}

%---------------------------------------------------------------------
\subsection{Asymptotic BER in Downlink}
%---------------------------------------------------------------------
Analysing the received signal of the $k$th user of the $\ell$th cell \eqref{eq:rx_kl_mf_simp}, we can also obtain some information about the bit-error probability: 
\begin{align}\label{eq:rx_kl_mf_simp2}
r_{k \ell} =& \sqrt{N \gamma_{k \ell}} \left(\frac{\sqrt{\phi_{k \ell}} \beta_{\ell k \ell} x_{k \ell}}{\alpha_{k \ell}} + \sum_{\substack{l = 1 \\ l \neq \ell}}^{L} \frac{\sqrt{\phi_{k l}} \beta_{l k \ell} x_{k l}}{\alpha_{k l}}\right) + n_{k \ell}.
\end{align}
\normalsize
Indeed, the effect of AWGN is averaged out when $N \rightarrow \infty$. For notation simplicity, but with no loss of generality, we consider 4-QAM modulation. Thus, the probability of error for this user can be written as \eqref{eq:Pe_bit}, where ${\rm Pr} (\cdot)$ is the probability of an event. Hence, \eqref{eq:Pe_bit} can be simplified as \colk{\eqref{eq:Pe_bit_simp}}, since both terms in the sum have the same statistical behaviour. The errors will occur whenever the interfering signal that reaches the user is greater than its intended signal. 

%\normalsize
\begin{subequations}\label{eq:Pe_bit_all}
\begin{align}
{\rm Pe}_{k \ell} = & \frac{1}{2} {\rm Pr} \left(\Re \left\{ \frac{\sqrt{\phi_{k \ell}} \beta_{\ell k \ell} x_{k \ell}}{\alpha_{k \ell}} \right\} < 
\Re \left\{ \sum_{\substack{l = 1 \\ l \neq \ell}}^{L} \frac{\sqrt{\phi_{k l}} \beta_{l k \ell} x_{k l}}{\alpha_{k l}}\right\}\right) + \notag \\
&\hspace{2cm} + \frac{1}{2} {\rm Pr}  \left(\Im \left\{ \frac{\sqrt{\phi_{k \ell}} \beta_{\ell k \ell} x_{k \ell}}{\alpha_{k \ell}} \right\} < 
\Im \left\{ \sum_{\substack{l = 1 \\ l \neq \ell}}^{L} \frac{\sqrt{\phi_{k l}} \beta_{l k \ell} x_{k l}}{\alpha_{k l}}\right\}\right), \label{eq:Pe_bit}\\ 
= \, & \frac{1}{2} {\rm Pr}  \left( \frac{\sqrt{\phi_{k \ell}} \beta_{\ell k \ell} \Re \left\{x_{k \ell} \right\}}{\alpha_{k \ell}} < 
\sum_{\substack{l = 1 \\ l \neq \ell}}^{L} \frac{\sqrt{\phi_{k l}} \beta_{l k \ell} \Re \left\{ x_{k l}\right\}}{\alpha_{k l}}\right) + \notag\\
&\hspace{2cm} + 
 \frac{1}{2} {\rm Pr}  \left(\frac{\sqrt{\phi_{k \ell}} \beta_{\ell k \ell} \Im \left\{ x_{k \ell} \right\}}{\alpha_{k \ell}} < 
\sum_{\substack{l = 1 \\ l \neq \ell}}^{L} \frac{\sqrt{\phi_{k l}} \beta_{l k \ell} \Im \left\{ x_{k l}\right\}}{\alpha_{k l}}\right), \nonumber\\
= \,& {\rm Pr}  \left( \frac{\sqrt{\phi_{k \ell}} \beta_{\ell k \ell} \Re \left\{x_{k \ell} \right\}}{\alpha_{k \ell}} < 
\sum_{\substack{l = 1 \\ l \neq \ell}}^{L} \frac{\sqrt{\phi_{k l}} \beta_{l k \ell} \Re \left\{ x_{k l}\right\}}{\alpha_{k l}}\right). \label{eq:Pe_bit_simp}
\end{align}
\end{subequations}
\normalsize

In order to determine the exact value of the probability in \eqref{eq:Pe_bit_simp}, we must analyse every possible combination of interfering signals. Thus, the result can be written as
\begin{eqnarray}\label{eq:Pe_bit_final}
{\rm Pe}_{k \ell} = \frac{1}{2^{L-1}}\sum_{j=1}^{2^{L-1}} {\rm u}\left[\left( \sum_{\substack{l = 1 \\ l \neq \ell}}^{L} \frac{\sqrt{\phi_{k l}} \beta_{l k \ell} b_{j l}}{\alpha_{k l}}\right) - \frac{\sqrt{\phi_{k \ell}} \beta_{\ell k \ell}}{\alpha_{k \ell}} \right]
\end{eqnarray}
\normalsize
where $b_{j l}$ is the $j,l$-th element of the $2^{L-1}\times L$ matrix ${\bf B} = [{\bf B}'_{1:\ell-1}, {\bf 1}_{2^{L-1}}, {\bf B}'_{\ell:L-1}]$, in which ${\bf B}'$ contains every possible combination of $\{\pm 1\}^{L-1}$. \colk{Although we restricted our investigation for 4-QAM modulation, similar analysis can be conducted for $M>4$ by appropriately defining the decision bounds for $r_{k \ell}$ in \eqref{eq:rx_kl_mf_simp2}.}

The expression in \eqref{eq:Pe_bit_final} gives the exact BER of the $k$th user of the $\ell$th cell, as a function of the powers and the long-term fading coefficients of users in adjacent cells sharing the same training sequence. Thus it can be adopted as a performance optimization metric, in the same way as defined in eq. \eqref{eq:ultimate_SINR}. {One can minimize it by varying users' transmit powers, and/or by {\it {changing} the assignment of pilot sequences to users}, as in the pilot allocation procedure}. 

%%%%%%%%%%%%%%%%%%%%%%%%%%%%%%%%%%%%%%%%%%%%%%
\section{Pilot Allocation Schemes}\label{sec:PA_Schemes}
%%%%%%%%%%%%%%%%%%%%%%%%%%%%%%%%%%%%%%%%%%%%%%%
Eq. \eqref{eq:rx_kl_mf_simp2} shows that the received signal for a given user in the downlink of a massive MIMO system presents interference from another users in adjacent cells that share the same pilot sequence. Besides, from this received signal in the limit of $N \rightarrow \infty$, the asymptotic expressions for SINR, eq. \eqref{eq:ultimate_SINR}, and  for BER,  eq. \eqref{eq:Pe_bit_final}, have been derived. At first glance, it may appear that the interference term in \eqref{eq:rx_kl_mf_simp2} does not depend on which user in $\ell$th cell is assigned the $k$th pilot sequence. However, reminding that $\alpha_{k l}^2 \cas \sum_{j=1}^{L} \gamma_{k j} \beta_{l k j} + \frac{1}{K}$, one can see it is not true. 

Thus, varying to which user is assigned the $k$th pilot sequence according its long-term fading coefficient can enhance the SINR, eq. \eqref{eq:ultimate_SINR}, and/or\footnote{Maximizing the SINR not necessarily minimizes the BER in the limit of $N \rightarrow \infty$, as can be seen from expressions \eqref{eq:ultimate_SINR} and \eqref{eq:Pe_bit_final}, and discussed in Sec. \ref{sec:results_PA}.}  decrease the probability of error, eq. \eqref{eq:Pe_bit_final}. This fact allows us the formulation of alternative optimization criteria, as described in the sequel. 

Initially, we define the matrix $\bf C$, of size $K!\times K$, containing every possible combination of pilot sequences to the users, \textit{i.e.}, $c_{i j}$ says that, in the $i$th combination, the $j$th pilot sequence is allocated to the $c_{i j}$th user. Then we define four pilot allocation {criteria} aiming to optimise the BER or {alternatively the} SINR figure{s} of metric. The four criteria are described in the following. 

%-------------------------------------------------
\subsection{MinBER-based Pilot Allocation Metric}
%-------------------------------------------------
In the first pilot allocation scheme, we search the best pilot distribution in the sense of minimising the mean BER among users of the $\ell$th cell, leading to the MinBER pilot allocation scheme:
%\small
\begin{equation}\label{eq:MinBER_PA}
i_{\textsc{mb}} = \arg \min_{i} \frac{1}{K} \sum_{k=1}^{K} {\rm Pe}_{c_{i k} \ell},
\end{equation}
in which
%\small
\begin{align}\label{eq:BER_pil_k}
& {\rm Pe}_{c_{i k} \ell} = \frac{1}{2^{L-1}} \times \sum_{j=1}^{2^{L-1}} {\rm u}\left[\left( \sum_{\substack{l = 1 \\ l \neq \ell}}^{L} \frac{\sqrt{\phi_{k l}} \beta_{l k \ell} b_{j l}}{\alpha^{(i)}_{k l}}\right) - \frac{\sqrt{\phi_{c_{i k} \ell}} \beta_{\ell c_{i k} \ell}}{\alpha^{(i)}_{c_{i k} \ell}} \right] \notag
\end{align}
\normalsize
corresponds to the BER of the $c_{i k}$th user of $\ell$th cell when the $k$th pilot sequence is assigned to him. Note that the superscript in $\alpha^{(i)}_{k l}$ and $\alpha^{(i)}_{c_{i k} \ell}$ evidences that these terms depend on the $i$th pilot distribution, since $(\alpha^{(i)}_{k l})^2 = \gamma_{c_{i k} \ell} \beta_{l c_{i k} \ell} + \sum_{\substack{j = 1 \\ j \neq \ell}}^{L} \gamma_{k j} \beta_{l k j} + \frac{1}{K}$, and $(\alpha^{(i)}_{c_{i k} \ell})^2 = \gamma_{c_{i k} \ell} \beta_{\ell c_{i k} \ell} + \sum_{\substack{j = 1 \\ j \neq \ell}}^{L} \gamma_{k j} \beta_{\ell k j} + \frac{1}{K}$, where we have put in evidence the terms related to the $c_{i k}$th user of the $\ell$th cell. 

{Note that} the pilot allocation procedure is evaluated in a decentralized way; therefore, the assignment of pilots can be modified only for users of the $\ell$th cell when it is carrying out this procedure. Although the strictly optimal solution would test every possible pilot combination among the $K\cdot L$ users, its complexity would be prohibitive. Thus, the solution obtained in the decentralized way is preferable, {and it can be shown that it converges to a Nash equilibrium after performing some times by each cell. Note that this problem can be viewed as a {\it finite potential game}, since: {\it a}) it exists a global potential function that maps every strategy to some real value according its efficiency; {\it b}) the set of strategies is of finite dimension \cite{Nisan07}. In this case, each cell is a player, the potential function would be the average BER of the whole system in that subcarrier, \textit{i.e.}, the average of \eqref{eq:Pe_bit_final} evaluated for the $K\cdot L$ users, and the strategy is the pilot allocation in that cell. Since each player chooses its strategy following a selfish best response dynamics, the convergence of the game to a Nash equilibrium is assured in \cite[Theorem 19.12]{Nisan07}, \cite[Proposition 2.2]{Voorneveld00}. Besides, the decentralized solution can achieve appreciable gains in performance, as demonstrated} {by numerical results in Section \ref{sec:results}.}

%-------------------------------------------------
\subsection{MaxSINR-based Pilot Allocation Metric}
%-------------------------------------------------
In the same way, in the second pilot allocation scheme, we define the pilot distribution that maximizes the mean SINR in the downlink of the $\ell$th cell, namely MaxSINR pilot allocation scheme:
\begin{equation}\label{eq:MaxSINR_PA}
i_{\textsc{ms}} = \arg \max_{i} \frac{1}{K} \sum_{k=1}^{K} \text{SINR}^\textsc{dl}_{c_{i k} \ell},
\end{equation}
in which
\begin{equation}\label{eq:SINR_pil_k}
\text{SINR}^\textsc{dl}_{c_{i k} \ell} = \frac{\phi_{c_{i k} \ell} \beta^2_{\ell c_{i k} \ell}/(\alpha^{(i)}_{c_{i k} \ell})^2}{\sum_{\substack{j = 1 \\ j \neq \ell}}^L  \phi_{k j} \beta^2_{j k \ell}/(\alpha^{(i)}_{k j})^2}
\end{equation}
is the downlink SINR of the $c_{i k}$th user of $\ell$th cell when the $k$th pilot sequence is assigned to him.

%-------------------------------------------------
\subsection{MiniMaxBER-based Pilot Allocation Metric}
%-------------------------------------------------
Both {previous performance} optimization schemes, {given respectively by eq.} \eqref{eq:MinBER_PA} and \eqref{eq:SINR_pil_k}, find the pilot distribution by optimizing the mean value of some performance criterion. However, the "average" approach may be not completely adequate in modern communications systems, since it may lead to a great improvement in performance for a few users, while providing low quality of service (QoS) to those users poorly located, typically in the edge of the cell. Hence, we also look for pilot allocation schemes that ensure improvement in QoS for every user within the $\ell$th cell. The MinimaxBER pilot allocation scheme {in multi-celular massive MIMO can be} defined as:
\begin{equation}\label{eq:MinimaxBER_PA}
i_{\textsc{mmb}} = \arg \min_{i} \max_{k} {\rm Pe}_{c_{i k} \ell},
\end{equation}
which minimizes the worst BER within the cell.

%-------------------------------------------------
\subsection{MaxMinSINR-based Pilot Allocation Metric}
%-------------------------------------------------
On the other hand, the MaxminSINR pilot allocation criterion constitutes an alternative way to optimally allocate pilots in multi-celular massive MIMO systems. The MaxminSINR pilot allocation scheme can be defined from the following optimization problem:
\begin{equation}\label{eq:MaxminSINR_PA}
i_{\textsc{mms}} = \arg \max_{i} \min_{k} \text{SINR}^\textsc{dl}_{c_{i k} \ell},
\end{equation}
which \colk{finds the pilots' set} that maximizes the lowest SINR among the users of the cell.

{In this paper, our objective consists in  {guaranteeing} a target QoS for the majority of the users. Thus, the MaxminSINR approach is the most suitable criterion of pilot allocation. {Hence, we} have considered this technique in conjucntion with power {allocation procedure} in the numerical results of Sec. \ref{sec:results_PA}.}

%-----------------------------------------------------------
\subsection{Pilot Allocation Algorithm and Its Complexity}
%-----------------------------------------------------------
In the analysis of the pilot allocation strategies for multi-cellular massive MIMO we have assumed the asymptotic condition, i.e, when $N \rightarrow  \infty$. Hence, the complexity of implementation of these pilot allocation algorithms will be independent of the number of antennas $N$. Algorithm \ref{alg:PA_proced} describes the general pilot allocation procedure, defining its inputs, outputs, and main steps. After its computation, the $\ell$th cell assigns the $k$th pilot sequence to the $c_{i_{opt}k}$-th user. Note that each cell should be able to find the optimal pilot combination set among its covered users following one of these four criteria given respectively by eq. \eqref{eq:MinBER_PA}, \eqref{eq:MaxSINR_PA}, \eqref{eq:MinimaxBER_PA}, \eqref{eq:MaxminSINR_PA}, in a decentralized way, reducing the overall computational complexity of the massive MIMO system.

\begin{algorithm}
\caption{Pilot Allocation Procedure}\label{algo:power_aloc}
\begin{flushleft}
Input: {$\mathcal{B}_{jl}$, ${\bm \Phi}_j$, ${\bm \Gamma}_j$, $\forall j,l =\nolinebreak 1,2,\ldots L$.}%\\
\end{flushleft}
\label{alg:PA_proced}
\begin{algorithmic}[1]
\STATE{Generate matrix $\bf C$, of size $K!\times K$;}
\FOR{each combination $i = 1, 2, \ldots, K!$}
	\FOR{each pilot sequence $k = 1, 2, \ldots, K$}
		\STATE{Evaluate \\{$\alpha^{(i)}_{c_{i k} \ell} = \sqrt{\gamma_{c_{i k} \ell} \beta_{\ell c_{i k} \ell} + \sum_{\substack{j = 1 \\ j \neq \ell}}^{L} \gamma_{k j} \beta_{\ell k j} + \frac{1}{K}}$};}\label{alpha_ev1}
		\FOR{each cell $l = 1, 2, \ldots, L$, $l \neq \ell$}
			\STATE{Evaluate \\ {$\alpha^{(i)}_{k l} = \sqrt{\gamma_{c_{i k} \ell} \beta_{l c_{i k} \ell} + \sum_{\substack{j = 1 \\ j \neq \ell}}^{L} \gamma_{k j} \beta_{l k j} + \frac{1}{K}}$};}\label{alpha_ev2}
		\ENDFOR
	\ENDFOR
\ENDFOR
\STATE{Find $i_{opt} \in i = 1, 2, \ldots, K!$, corresponding to the optimal combination in $\bf C$ according some metric: \eqref{eq:MinBER_PA}, \eqref{eq:MaxSINR_PA}, \eqref{eq:MinimaxBER_PA}, \eqref{eq:MaxminSINR_PA};}\label{line_p}
\end{algorithmic}
Output: $i_{opt}$.
\end{algorithm}

Indeed, the computational complexity for the SINR-based pilot allocation procedures  results $\mathcal{O}(K!\cdot K \cdot L^2)$, since operation of lines \ref{alpha_ev1} and \ref{alpha_ev2} demands $2L + 1$ {floating point operations (\emph{flops})} each one, and will be evaluated $K!\cdot K \cdot L$ times. On the other hand, the BER-based pilot allocation schemes result in a computational complexity of $\mathcal{O}(K!\cdot K \cdot L \cdot 2^{L-1})$, since evaluation of Eq. \eqref{eq:MinBER_PA} or \eqref{eq:MinimaxBER_PA} in line \ref{line_p} becomes prevalent, \textit{i.e.} of $\mathcal{O}(K \cdot L \cdot 2^{L-1})$, and should be evaluated for the $K!$ pilot combinations. Such complexities may appear excessive. However, bearing in mind that $K$ must assume low values in practical scenarios\footnote{$K$ represents the length of the training sequences, which is equal to the number of users sharing one of $N_{\textrm{smooth}}$ subcarriers in each coherence band. \colk{$N_{\textrm{smooth}}$, as given in \eqref{eq:Nsmooth}, is desired to be high for an efficient OFDM communication; for example, $N_{\textrm{smooth}} = 14$ for $\Delta t_{CP} = 7\%$.} On the other hand, $K$ is limited by the coherence time, due to the user's mobility, by the efficiency of the TDD scheme, which cannot spend much time with pilots, and by the subcarrier spacing. For example, in \cite{Marzetta10}, for a cell serving 42 users, $K = 3$ and $N_{\textrm{smooth}} = 14$.}, as well as the number $L$ of cells within a cluster, one can conclude that the complexity of pilot allocation procedures is not prohibitive. \colk{As described in \cite{Marzetta10}, there is no appeal to consider higher values of $K$ in practical mobile TDD massive MIMO scenarios, since great part of the coherence time interval would be spent acquiring CSI from the moving terminals.} Besides, once the optimization is complete, it remains valid for a relatively large time-interval, since the scheme depends only on the transmit powers and long-term fading coefficients of the users. Even for specific scenarios in which $K$ may assume higher values, \colk{\textit{i.e.}, with reduced mobility such as pedestrian scenarios}, the exhaustive search approach can be replaced by some low-complexity heuristic method, and the optimization remains valid.

%%%%%%%%%%%%%%%%%%%%%%%%%%%%%%%%%%%%%%%%%%%%%%
\section{{Power Control Algorithm}}\label{sec:PowC}
%%%%%%%%%%%%%%%%%%%%%%%%%%%%%%%%%%%%%%%%%%%%%%%

{With the purpose of serving the users with a target SINR, the target-SIR-tracking algorithm measures the interference seen by each user, and  assigns to him the exact power to reach the target SINR, unless if this power exceeds the maximum power available. In this case, the maximum power is allocated for this user. It is shown in \cite{Rasti11} that this power allocation procedure is not the most suitable, since assigning the maximum power for the users with poor channel conditions causes an excessive interference for the other users, and waste energy because this user may remain with a low SINR. Being $\widehat{\zeta}^\textsc{dl}_{k \ell}$ the target downlink SINR for the $k$-th user of the $\ell$-th cell, and $\overline{\phi}_{k \ell}$ the maximum transmit power that can be assigned to him, the target-SIR-tracking algorithm updates power at the $i$-th iteration according to}
{
\begin{equation}\label{eq:TPC_pow}
{\phi}_{k \ell}(i) = \min \left[ \widehat{\zeta}^\textsc{dl}_{k \ell} \mathcal{I}_{k \ell}(i), \overline{\phi}_{k \ell} \right],
\end{equation}}
{in which {$\mathcal{I}_{k \ell}(i) = \frac{\phi_{k \ell}(i-1)}{\text{SINR}^\textsc{dl}_{k \ell}(i-1)}$} is the interference seen by this user divided by $\beta^2_{\ell k \ell}/\alpha^2_{k \ell}$. By contrast, the power control algorithm proposed in \cite{Rasti11} updates users' powers as}
{
\begin{eqnarray}\label{eq:OPC_pow}
{\phi}_{k \ell}(i) = \left\{\begin{matrix}
\widehat{\zeta}^\textsc{dl}_{k \ell} \mathcal{I}_{k \ell}(i) & {\rm if} \quad \mathcal{I}_{k \ell}(i)\leq \frac{\overline{\phi}_{k \ell}}{\widehat{\zeta}^\textsc{dl}_{k \ell}},\\ 
\frac{\overline{\phi}_{k \ell}^2}{\widehat{\zeta}^\textsc{dl}_{k \ell} \mathcal{I}_{k \ell}(i)} & {\rm otherwise.}
	\end{matrix}\right.
\end{eqnarray}}

{One can see that the method of \cite{Rasti11} assigns power to the users in the same way as the target-SIR-tracking algorithm if the target SINR can be achieved for the user at that iteration. Otherwise, instead of allocating him the maximum power, it allocates a transmit power inversely proportional to that required for achieve the target. Thus, besides of saving energy relative to users that cannot reach the target SINR, the interference irradiated to other users also decreases.}

{When deploying the power control algorithm of \cite{Rasti11}, the target SINR parameter should be carefully chosen. If a somewhat lower value is adopted, the algorithm saves energy by delivering just the target SINR to the users, taking low advantage of the resources and providing poor performance for the system. If {an} excessive target SINR is considered, many poor located users will have their powers gradually turned off in order to provide the desired performance for the other users. Hence, in this paper, the target SINR was chosen in each scenario by finding the value that achieves the higher throughput for 95\% of the users, by means of numerical simulations.}

%%%%%%%%%%%%%%%%%%%%%%%%%%%%%%%%%%%%%%%%%%%%%%%
\section{Numerical Results}\label{sec:results}
%%%%%%%%%%%%%%%%%%%%%%%%%%%%%%%%%%%%%%%%%%%%%%%
Aiming to demonstrate the effectiveness of the proposed pilot allocation method {combined with power control} {strategies} for multi-cellular massive MIMO systems, we provide in this Section performance results for both BER and SINR downlink metrics. The asymptotic condition ($N\rightarrow \infty$) for the number of BS antennas has been assumed, except for the convergence analysis for increasing $N$ depicted in Fig \ref{fig:conv_perf}. We have adopted a multi-cell scenario with hexagonal cells of radius 1600m, where $K=4$ users are uniformly distributed in its interior, except in a circle of 100m radius around the cell centered BS. Besides, only the first ring of interfering cells has been considered, both for frequency reuse factors (RF) of one and three. We have assumed a similar TDD protocol of that in \cite{Fernandes13}, in which the coherence interval is composed of 11 symbol periods: 4 for sending uplink training sequences, 1 for processing, 4 and 2 for downlink and uplink data transmission, respectively. As discussed in \cite{Marzetta06_Conf}, in order to maximize the net throughput for a TDD protocol, it is beneficial to dedicate the same time with pilots and data transmissions. If more time is spent with pilots, more users can be served, but its rates decrease substantially due to the excessive overhead, and vice-versa. The system uses a carrier frequency of 1.9 GHz and a frequency band of 20 MHz. 

Furthermore, the coherence time of 500 microseconds has been adopted, which could accommodate any terminal moving slower than 80 meters/second (associating the coherence time with the interval required by a terminal to move no more than 1/4 wavelength \cite[Sec.VII-D]{Marzetta10}). The log-normal shadowing has been modelled with a standard deviation of 8dB, and the path loss term $d_{\ell k j}^{-\lambda}$ with decay exponent equal to $\lambda = 3.8$, and $d_{\ell k j}$ denoting the distance between the $\ell$th BS to $k$th mobile user of $j$th cell. Besides, we have considered 4-QAM modulation, {and an equal uplink training power allocation with signal-to-noise ratio (SNR) of 10 dB for all users}. \colk{The constraint of maximum transmit power available at BS is satisfied by the precoding schemes in our formulation, as represented in expressions \eqref{eq:MF_prec} and \eqref{eq:ZF_prec}, in which the precoding vectors are normalized.} It is important to note that the numerical results in this Section (except part of Fig. \ref{fig:conv_perf}) were obtained from the analytical expressions derived in Sec. \ref{sec:Massive_MIMO}, averaged from the evaluation of at least $10^5$ independent trials for the user's location.

{A downlink transmit power relative to a SNR of 10dB is equally assigned to the users when the equal power allocation policy is assumed. On the other hand, when deploying {interference-based} power control, the power assignment described by \eqref{eq:OPC_pow} is carried out, {within} 10 iterations, that was demonstrated in \cite{Rasti11} to be sufficient for convergence. The maximum transmit power for each user is {such} that to achieve a mean SNR of 10dB, while the target SINR for each scenario is the one that increases the 95\%-likely rate for the users. By simulations, we have found the values of $\widehat{\zeta}^\textsc{dl}$ (the same for all users) shown in Tables \ref{tab:res_rf1&rf3}, which we have adopted for the power control algorithm for generating results of Section \ref{sec:results_PA}.}

\begin{figure}[!htbp]
\centering
{\small a) RF = 1 \hspace{45mm} b) RF = 3}\\
\includegraphics[width=.65\textwidth]{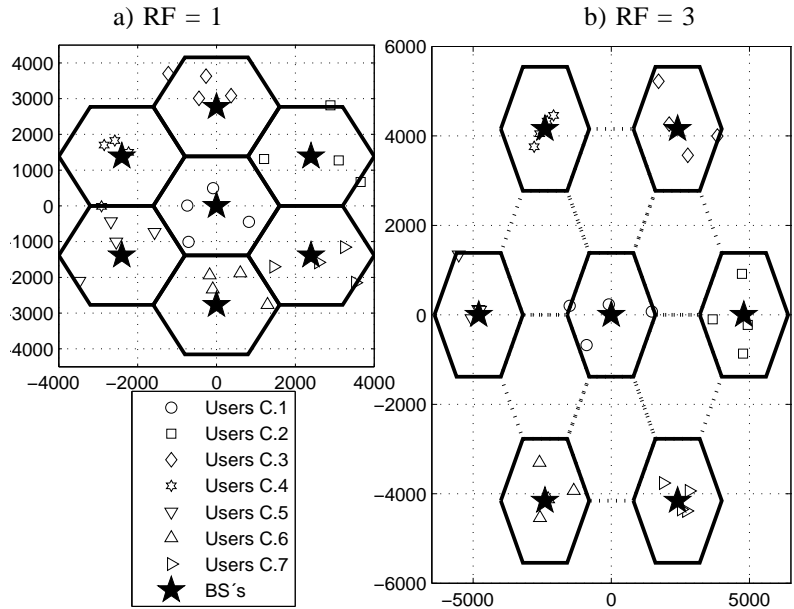}
\caption{Single spatial realization for both investigated multi-cell scenarios, with $K = 4$ mobile terminais.}
\label{fig:mc_spatial_dist}
\end{figure}

Figure \ref{fig:mc_spatial_dist} depicts a single realization of the multi-cell scenario adopted in our numerical simulations, for frequency reuse factors of one and three. Notice that for clarity purpose, only users sharing the same frequency band in the training stage, \textit{i.e.}, interfering with each other, have been represented. Indeed, one can see that interfering users are much closer with smaller reuse factors. In our numerical results presented in the sequel, only the performance metrics of users positioned inside the central cell were computed, since these users experience a more realistic condition of interference.

%---------------------------------------------------------------------
\subsection{Performance Convergence of Precoding Techniques}\label{sec:results_conv}
\begin{figure}[!htbp]
\centering
{\small a) BER \hspace{45mm} b) SINR}
\includegraphics[width=.85\textwidth]{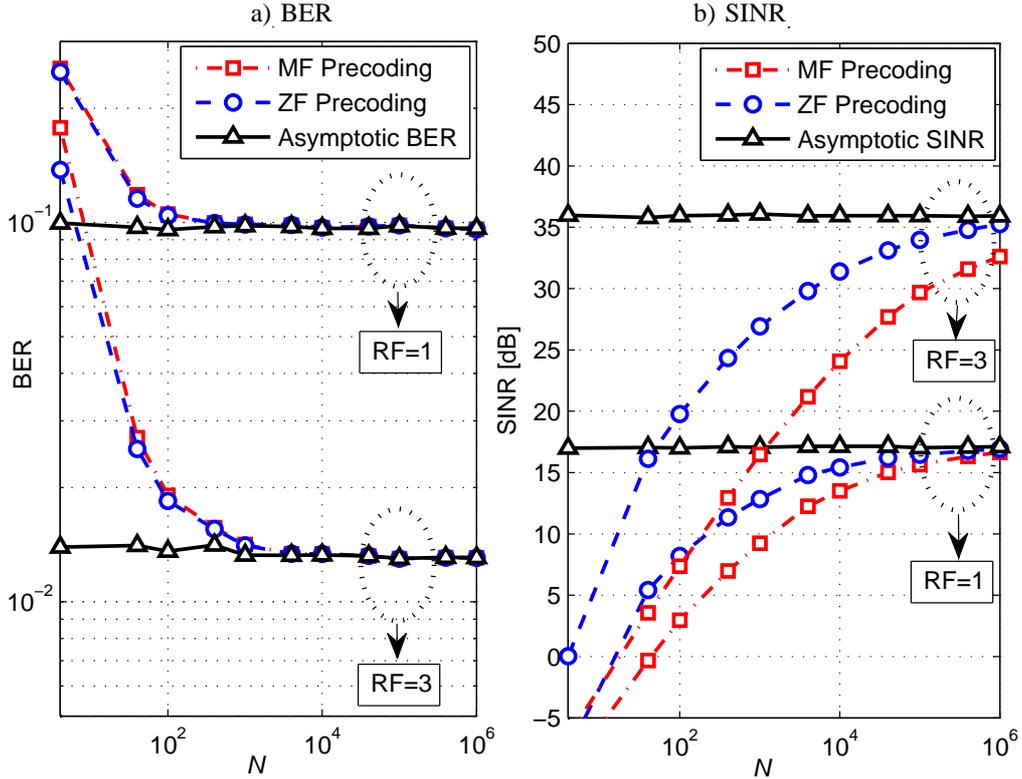}
\vspace{-1mm}
\caption{Asymptotic convergences of MF and ZF precoding techniques to the performance bounds, under reuse factors of one and three, with increasing $N$.}
\label{fig:conv_perf}
\end{figure}

%---------------------------------------------------------------------
Considering both MF and ZF precoding techniques, Fig. \ref{fig:conv_perf} depicts the asymptotic convergence\footnote{Notice that  in Fig. \ref{fig:conv_perf} simulation and analysis results have been compared, since performances of the techniques for increasing $N$ are computed with independent realizations of small-scale fading, AWGN, and long-term fading, and they converge to the analytical bounds dependent only on the long-term fading.} (as the number of BS antennas increases) for both BER and SINR performance metrics to the bounds defined in \eqref{eq:Pe_bit_final} and \eqref{eq:ultimate_SINR}, respectively. The curves present mean values of each performance metric, taken among the users of the cell, {under equal power assignment policy}. One can note that the SINR of the ZF precoding scheme indeed converges to the same bound of eq. \eqref{eq:ultimate_SINR}, which was derived in \cite{Fernandes13} as the asymptotic SINR of MF beamforming. This occurs since we have considered the constraint of maximum transmit power available at BS, as opposed to \cite{Rusek13}.  These numerical results also show that MF needs at least one order of magnitude more BS antennas than ZF to reach that bound. Furthermore, the performance of both schemes are also analysed from the perspective of BER, validating eq. \eqref{eq:Pe_bit_final} as the asymptotic BER that such techniques are able to achieve when $N \rightarrow \infty$. Indeed, in terms of BER, the performances of both techniques rapidly approach the asymptotic limit, being necessary $\approx 10^4$ BS antennas for both precoding techniques reaching the bound. 

%---------------------------------------------------------------------
\subsection{Performance of Pilot Allocation Scheme {with Power Control}}\label{sec:results_PA}
%---------------------------------------------------------------------
{In this subsection, we investigate the performance of the MaxminSINR pilot allocation scheme proposed in Sec. \ref{sec:PA_Schemes}, with and without power control, in terms of mean values, as well as in terms of distribution among users. The simulation results presented here were averaged over 100,000 spatial realisations. }

\begin{figure}[!htbp]
\centering
{\small a) RF = 1 \hspace{45mm} b) RF = 3}
\includegraphics[width=.75\textwidth]{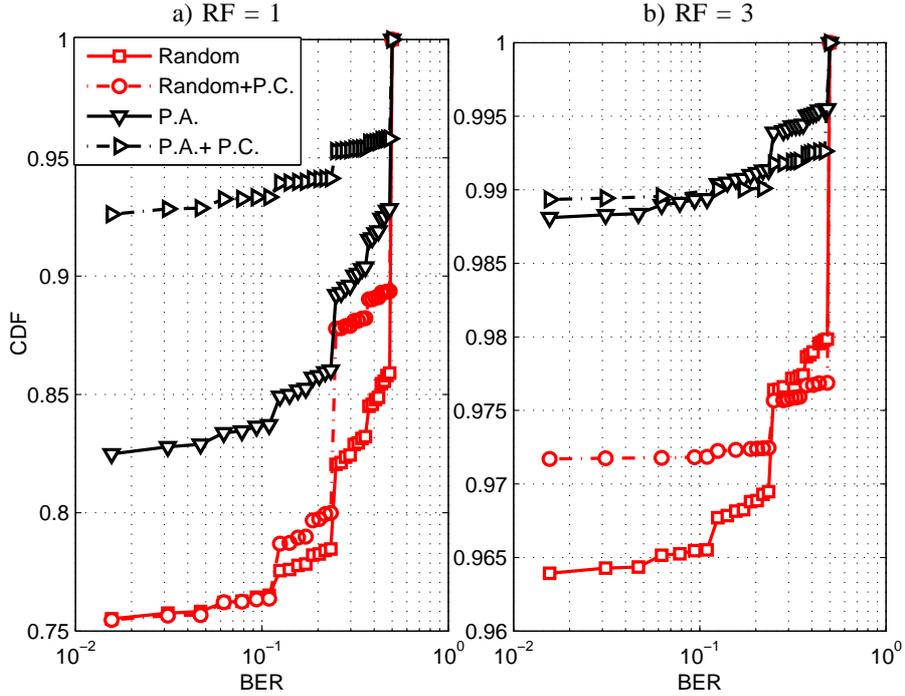}
\vspace{-.7mm}
\caption{Cumulative distribution function for the BER of the users, {when MaxminSINR pilot allocation is combined to power control.}}
\label{fig:CDF_BER}
\end{figure}

{Figure \ref{fig:CDF_BER} shows the cumulative distribution function (CDF) as a function of the BER of the users, regarding {the} MaxminSINR pilot allocation {combined to the} power control technique. For reference of comparison, it is also depicted the very large MIMO performance with no optimisation in the distribution of pilot sequences, \textit{i.e.}, with random allocation strategy.} An interesting behaviour on the BER distribution among users in the massive MIMO system can be observed from these numerical results. One can note that a significant portion of users communicates to BS with no errors, \textit{i.e.}, BER = 0. This occurs because, for these users, even the strongest interference that can reach them is lower than their intended signal, and thus the probability of error is null. On the other hand, the other small portion of users, that are not free of errors, presents excessive values of BER. This disparity becomes more noticeable for unitary frequency reuse factor, in which the portion of users that presents excessive {bit} error rates is $\approx 10\%$, while for reuse factor of three it is about $1\%$ for a BER $\geq 10\%$.

Furthermore, it shows that the pilot allocation scheme combined with power control is able to significantly decrease the fraction of users with excessive BER's. As shown in Tables \ref{tab:res_rf1&rf3}, the fraction of users with BER $\geq 0.1$ reduces from 23.63$\%$ to 6.65$\%$, for frequency reuse factor of one, and from 3.47$\%$ to 1.01$\%$, for frequency reuse factor of three, when deploying the MaxminSINR approach in conjunction with power control. 

\begin{figure}[!htbp]
\centering
{\small a) RF = 1 \hspace{45mm} b) RF = 3}
\includegraphics[width=.75\textwidth]{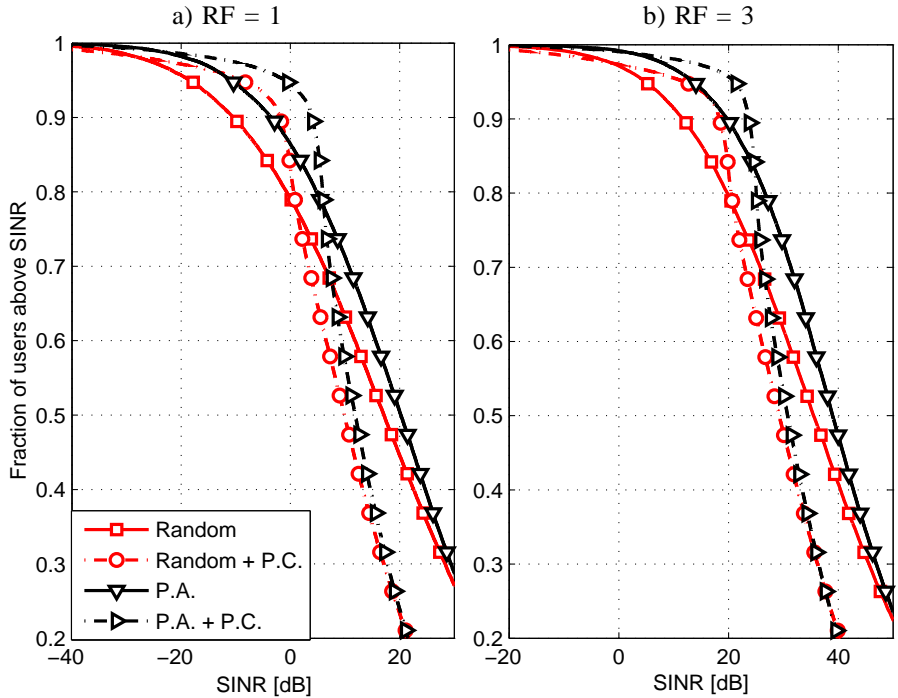}
\vspace{-.6mm}
\caption{Fraction of users above a given SINR {when MaxminSINR pilot allocation is combined to power control} {strategy.}}
\label{fig:invCDF_SINR}
\end{figure}

Figure \ref{fig:invCDF_SINR} shows the fraction of users above a given SINR, for frequency reuse factors of one and three. It can be seen that increasing the frequency reuse factor has the effect of significantly improving the SINR of the users, as if the curve was shifted right $\approx 22$dB, without noticeable changes on its format and slope. One can see that the application of the power control algorithm has the effect of making the fall of the SINR distribution curve more steep, decreasing the variation of SINR's among users, {increasing consequently the BER performance and the respective SINR. Moreover, if the} pilot allocation is jointly employed, it has the effect of {further increasing in} the SINR in which the fall occurs. In summary, pilot allocation combined with power control benefits the less favourably located users.
%\vspace{-3mm}

\small
\begin{table}[!htbp]
\caption{{Performance of pilot allocation and power control for frequency-reuse factor of one {and three (RF$=1$ and RF$=3$)}.}}
\vspace{2mm}
\centering
\footnotesize
\begin{tabular}{|c|c|c|c|c|c|}
\hline
\bf PA      & Mean  & Users   & Users        & Mean   		&  $95\%$-likely \\
\bf Scheme  & BER   & BER=0   & BER$\geq$0.1 & user Rate  & user Rate      \\
\bf         & (\%)  & (\%)    & (\%)         & (Mbps) 		& (Mbps)         \\
\hline\hline
\multicolumn{6}{|c|}{{\bf Frequency-reuse factor  RF$=1$}}\\
\hline
Random & $9.84$ & $75.41$ & $23.63$ & $48.50$ & $0.1344$\\
\hline
Random + P.C. & $8.44$ & $75.47$ & $23.57$ & $30.89$ & $1.4610$\\
 ($\widehat{\zeta}^\textsc{dl}$ = 0dB) &&&&&\\
\hline
MaxminSINR & $6.17$ & $82.45$ & $16.28$ & $52.62$ & $0.7937$\\
\hline
MaxminSINR + P.C. & $2.73$ & $92.61$ & $6.65$ & $34.24$ & $6.7430$\\
($\widehat{\zeta}^\textsc{dl}$ = 6dB)  &&&&&\\
\hline\hline
\multicolumn{6}{|c|}{{\bf Frequency-reuse factor  RF$=3$}}\\
\hline
Random & $1.41$ & $96.33$ & $3.47$ & $29.07$ & $4.79$\\
\hline
Random + P.C. & $1.29$ & $97.17$ & $2.82$ & $24.39$ & $10.41$\\
($\widehat{\zeta}^\textsc{dl}$ = 20dB)  &&&&&\\
\hline
MaxminSINR & $0.39$ & $98.78$ & $1.09$ & $31.68$ & $11.15$\\
\hline
MaxminSINR + P.C. & $0.45$ & $98.94$ & $1.01$ & $25.95$ & $17.31$\\
($\widehat{\zeta}^\textsc{dl}$ = 25dB)  &&&&&\\
\hline
\end{tabular}
\label{tab:res_rf1&rf3}
\end{table}

\begin{figure}[!htbp]
\centering
{\small a) RF = 1 \hspace{45mm} b) RF = 3}
\includegraphics[width=.75\textwidth]{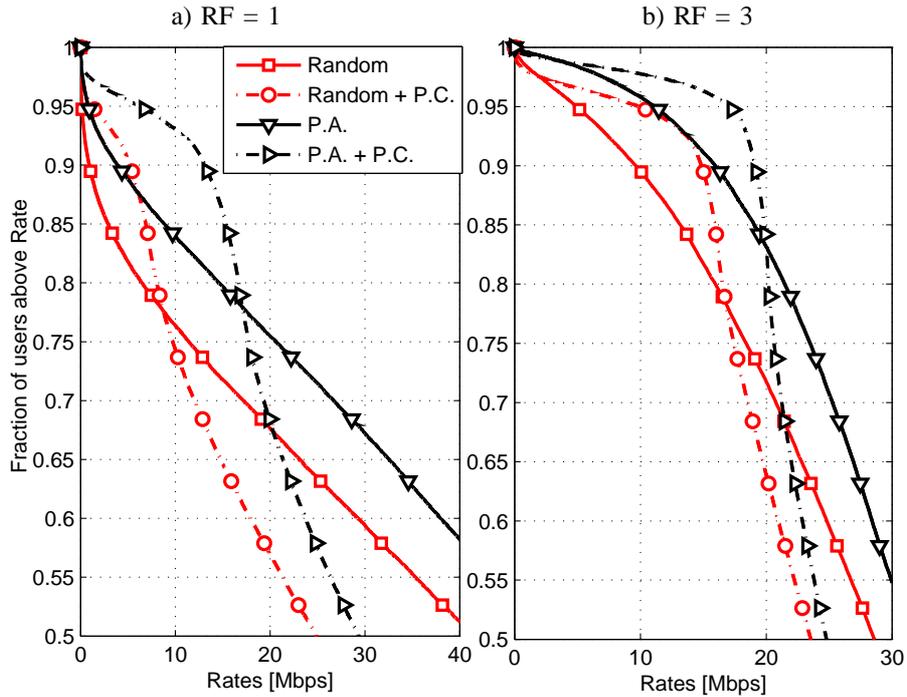}
\vspace{-.6mm}
\caption{Fraction of users above a given rate, {when MaxminSINR pilot allocation is combined to power control.}}
\label{fig:inv_CDF_Rate}
\end{figure}

\normalsize
Finally, Figure \ref{fig:inv_CDF_Rate} depicts the fraction of users above a given data rate, for frequency reuse factors of one and three, {regarding the {jointly} application of pilot allocation and power control} {procedures}. Notice that the downlink data rate $\mathcal{R}_{k \ell}$ {for} the $k$th user in the $\ell$th cell can be defined as:
\begin{equation}
\mathcal{R}_{k \ell} = \left( \frac{\textsc{bw}}{\textsc{rf}} \right)\left( \frac{\mathcal{D}}{\mathcal{T}} \right)\log_2\left(1 + \text{SINR}^\textsc{dl}_{k \ell}\right),
\end{equation}
where $\textsc{bw}$ is the system total bandwidth, $\textsc{rf}$ is the reuse factor, $\mathcal{D}$ is the number of symbol periods spent sending downlink data, and $\mathcal{T}$ is the total number of symbol periods within a channel coherence time.

Examining the curves, one can conclude that the slope of curves for reuse factor three is greater than the slope of unitary reuse factor curves. This fact means that the distribution for unitary reuse factor is much more irregular, unequal, in the sense that some users have very high rates while others have low QoS. On the other hand, for reuse factor of three, this distribution is much more uniform, guaranteeing simultaneously an improved QoS for much more users.
\vspace{-3mm}

As shown in Table \ref{tab:res_rf1&rf3} {for RF$=1$}, 95$\%$ of users {communicating under} equal power policy with rates greater than 0.1344Mbps with random pilot distribution, while when employing the MaxminSINR PA scheme the 95$\%$-likely rate per user increases to 0.7937Mbps. If power control is adopted, the proportional gain is slightly reduced, but the 95$\%$-likely rate per user of 6.743Mbps achieved is quite appreciable. Similar findings can be taken analysing the results for reuse factor of 3 in Table \ref{tab:res_rf1&rf3}, in which a formidable throughput of 17.31Mbps can be assured for 95\% of the users when combining pilot allocation and power control {strategies}. Note that the mean rate, however, decreases when enhancing the reuse factor, since the gain in SINR for the best located users does not offset the loss due to reduction in bandwidth, given the logarithmic increase of rate according SINR gains. Larger reuse factors are more beneficial for poor located users, since the logarithm is in its linear region, as discussed in \cite{Marzetta10}.

{Comparing  {results for RF$=1$ and  RF$=3$ in}  Table \ref{tab:res_rf1&rf3}, we note that the increase in the assured QoS due to power control for RF=1 that was from 134kbps to 1.461Mbps with no pilot allocation, turns to be from 793kbps to 6.743Mbps with MaxminSINR PA. The proportional gains slightly decrease with larger reuse factor, but an appreciable QoS can be assured for the users combining pilot allocation and power control. Besides, the portion of users communicating in the absence of errors increases from 75.41$\%$ to 92.61$\%$ for RF=1, and from 96.33\% to 98.94\% for RF=3. These benefits are achieved by simply assigning the pilot sequences to the users within the cell in a more efficient way, in conjunction with the application of power control algorithm, and remain valid whenever the long-term fading coefficients stay unchanged.}

%%%%%%%%%%%%%%%%%%%%%%%%%
\section{Conclusion}\label{sec:concl}
%%%%%%%%%%%%%%%%%%%%%%%%% 
In this work we have characterized the asymptotic performance of the massive MIMO system downlink under the point of view of the BER performance. Then, we derived the exact asymptotic expression of the BER of a given user, based on the long-term fading coefficients and the power levels of other users. In the same way as the asymptotic SINR expression found in \cite{Fernandes13}, the BER expression derived also depends only on the users in neighboring cells that reuse the same pilot sequence. 

{Furthermore, we have proposed efficient forms of assigning these pilots to the users within the cell, by optimizing several performance metrics. The significant gains achieved by the MaxminSINR pilot allocation technique in conjunction with power control were demonstrated numerically. For instance, we have showed that a gain of 50 times (0.1344 to 6.7430 Mbps) can be achieved for the downlink rate with unitary reuse factor combining both techniques, while the data rate is increased from 4.79 Mbps to 17.31 Mbps for reuse factor of three. For the last reuse factor, we showed that the massive MIMO system are able to operate with a 95$\%$-likely downlink rate of 17.31 Mbps, providing a communication free of errors for 98.94$\%$ of the users, guaranteeing the reliability of the system. }

{All of these benefits are achieved in a quite simple and expeditious way, by just knowing the powers and the long-term fading coefficients of users in adjacent cells, for each pilot sequence. Since these informations do not scale with the number of BS antennas, and remains constant within a long time and frequency interval, the implementation of the proposed pilot assignment method in conjuntion with power control algorithm for massive MIMO system is surely feasible.}

\section*{Acknowledgement}
This work was supported in part by the National Council for Scientific and Technological Development (CNPq) of Brazil under Grant 304066/2015-0, and in part by CAPES (scholarship), and by Londrina State University - Paraná State Government (UEL).

%
%
%\bibliographystyle{spmpsci}      % mathematics and physical sciences2
%\bibliography{la_mimo_refs}

\end{document}